\documentclass[a4paper,10pt]{article}
\usepackage{amssymb}
\usepackage{amsmath}
\usepackage{latexsym}
\usepackage{array}
\usepackage{graphicx}

\def\DIS{\displaystyle}

\usepackage{theorem}
\theoremstyle{break}
\newtheorem{Theorem}{Theorem}
\newtheorem{Proposition}{Proposition}

\def\C{{\mathbb C}}
\def\Z{{\mathbb Z}}

\begin{document}
\title{Conserved quantities and generalized solutions of the ultradiscrete KdV equation}
\author{Masataka Kanki${}^1$, Jun Mada${}^2$ and Tetsuji Tokihiro${}^1$ \\ \\
${}^1$ Graduate school of Mathematical Sciences, \\
      University of Tokyo, 3-8-1 Komaba, Tokyo 153-8914, Japan\\
${}^2$ College of Industrial Technology, \\
  Nihon University, 2-11-1 Shin-ei, Narashino, Chiba 275-8576, Japan}

\date{}

\maketitle

\begin{abstract}
We construct generalized solutions to the ultradiscrete KdV equation, including the so-called negative solition solutions.
The method is based on the ultradiscretization of soliton solutions to the discrete KdV equation with gauge transformation.
The conserved quantities of the ultradiscrete KdV equation are shown to be constructed in a similar way to those for the box-ball system.
\end{abstract}

\section{Introduction}
\label{sec1}

\begin{figure}[t]
 \begin{center}
 \includegraphics[width=.8\linewidth]{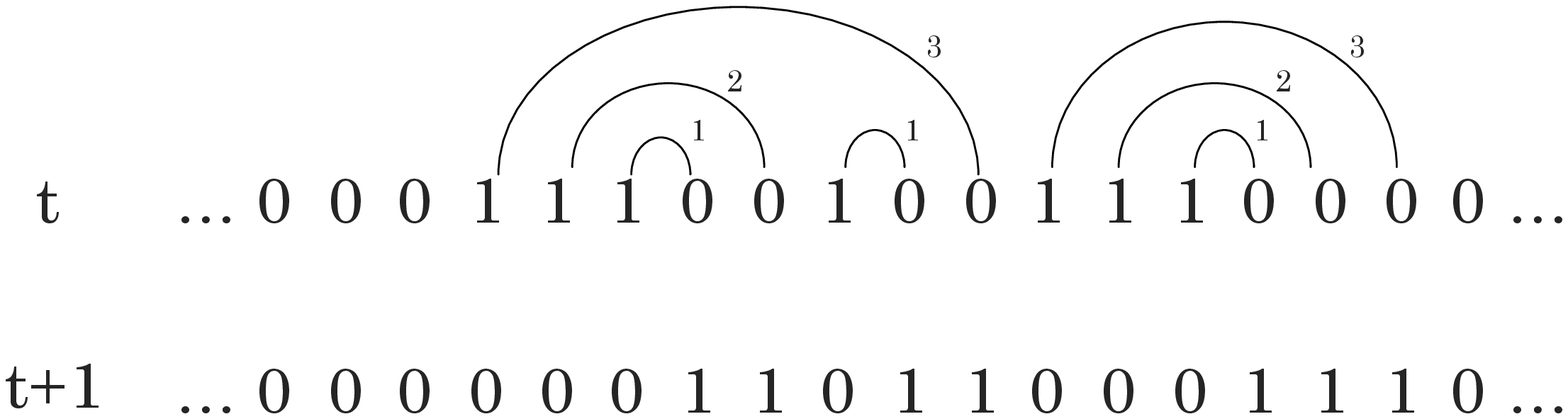} 
 \caption{The update rules of the BBS. A vacant box and a box with a ball are represented  by `0' and `1',
respectively. The conserved quantities are simultaneously determined. ($q_1=3,\ q_2=q_3=2$.)}
 \label{fig01}
 \end{center}
\end{figure}

\begin{figure}[t]
 \begin{center}
 \includegraphics[width=.6\linewidth]{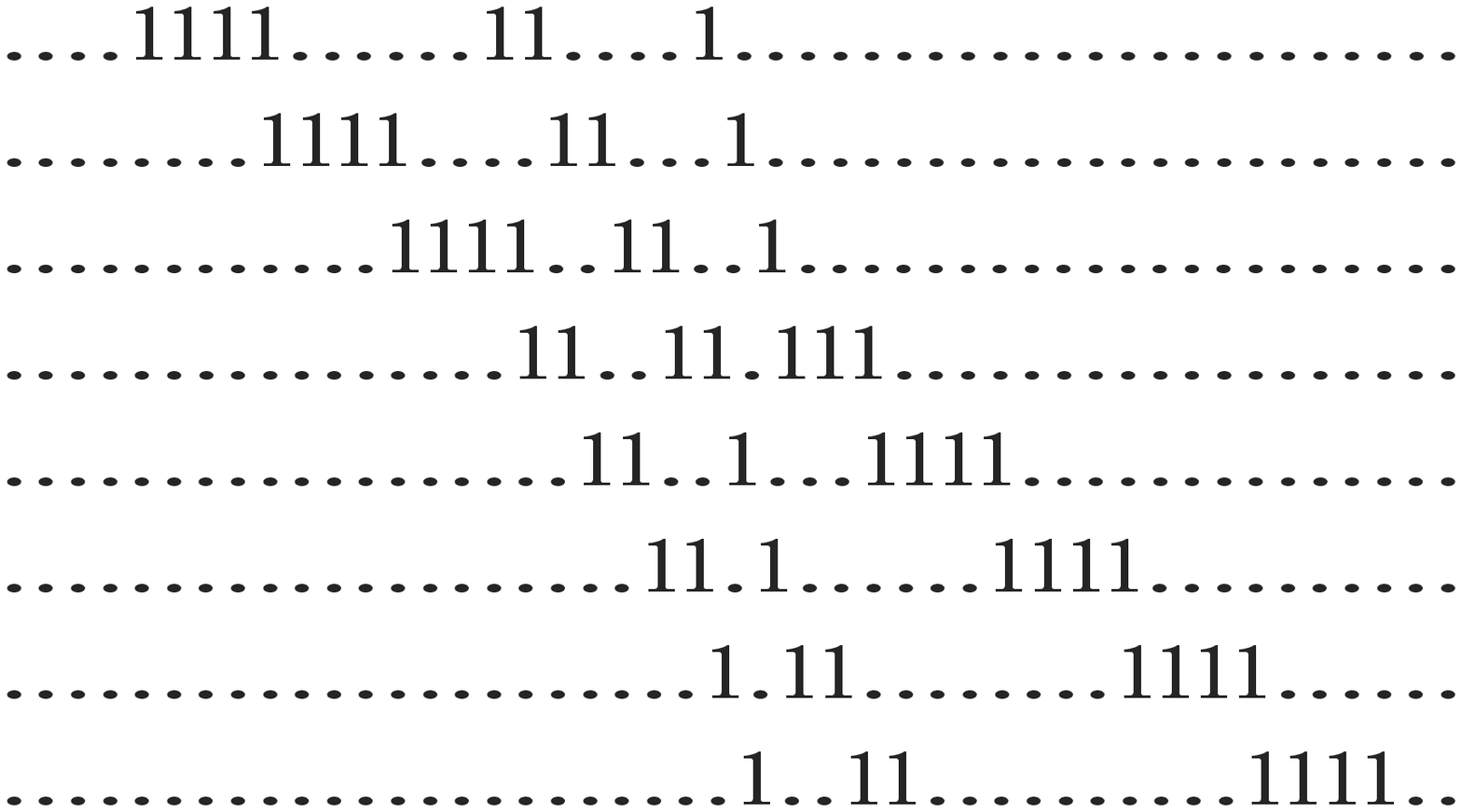} 
 \caption{A solitonic time evolution pattern of the BBS. The symbol `.' denotes `0'.}
 \label{fig02}
 \end{center}
\end{figure}

The ultradiscrete KdV equation:
\begin{equation}
U_n^{t+1}=\min \left[1-U_n^t, \sum_{k=-\infty}^{n-1} \left(U_k^t-U_k^{t+1}  \right) \right]
\label{UDKDV}
\end{equation}
was first introduced as a dynamical equation for the so-called box-ball system (BBS)\cite{TaS1990, Takahashi}.
Since \eqref{UDKDV} is closed on the binary values $0$ or $1$, 
if we regard $U_n^t$ as the number of balls in the $n$th box at time step $t$, and assume the boundary condition $\DIS \lim_{|n| \to \infty}U_n^t =0 $,
it describes the time evolution of balls in a one dimensional array of boxes.
The updating rule of the BBS according to \eqref{UDKDV} can be expressed as follows. (See figure~\ref{fig01}.)
\begin{itemize}
\item Find all the pairs of a ball and an adjacent vacant box to the right of it, and draw arclines from the ball to the vacant box for each of these pairs.
\item Neglecting the pairs connected by arclines, find all the pairs of a ball and an adjacent vacant box to the right of it and draw arclines in a similar way.
\item Repeat the above procedure until all the balls are connected to vacant boxes, then move all the balls into the vacant boxes connected by arclines.
\end{itemize}
An example of the time evolution pattern is shown in figure~\ref{fig02}.
The consecutive balls behave like solitons in the KdV equation.
Furthermore, when we denote by $q_1$ the number of arclines drawn in the first step of the updating rule, 
by $q_2$ one in the second step and so on, 
we obtain a nonincreasing integer sequence $(q_1,q_2,q_3, \ldots)$, which is a conserved quantity of the BBS 
in time\cite{ToTaS}.

The reason why the BBS has these solitonic properties is understood by the fact that \eqref{UDKDV} is obtained from the KdV equation
through a limiting procedure, ultradiscretization\cite{TTMS}.
The process involved in obtaining \eqref{UDKDV} from the KdV equation can be explained as follows.
The integrable discrete KdV equation is given in the bilinear form as
\begin{equation}
(1+\delta)\sigma_{n+1}^{t+1}\sigma_n^{t-1}=\delta\sigma_{n+1}^{t-1}\sigma_n^{t+1}+\sigma_n^t\sigma_{n+1}^t,
\label{DKdVbilinear}
\end{equation}
where $(n,t) \in \Z^2$ and $\delta$ is a discretization parameter\cite{HTI}.
By putting 
\begin{equation}
 u_n^t:=\frac{\sigma_n^t\sigma_{n+1}^{t-1}}{\sigma_{n+1}^t\sigma_n^{t-1}},\quad
v_n^t:=\frac{\sigma_n^{t+1}\sigma_{n}^{t-1}}{(\sigma_n^t)^2}, 
\label{SigmatoUandV}
\end{equation}
\eqref{DKdVbilinear} gives the simultaneous equations
\begin{align}
(1+\delta)\frac{1}{u_n^{t+1}}&=\delta u_n^t +\frac{1}{v_n^t}, 
\label{DKdVc1} \\
u_n^{t+1}v_{n+1}^t&=u_n^tv_n^t.
\label{DKdVc2}
\end{align}
When we impose the boundary condition $\displaystyle \lim_{|n| \to \infty} \sigma_n^t= \mbox{const.}$, 
then $\displaystyle \lim_{|n| \to \infty}v_n^t = \lim_{|n| \to \infty}u_n^t =1$
 from \eqref{DKdVc1} and \eqref{DKdVc2} we have
\begin{equation}
(1+\delta)\frac{1}{u_n^{t+1}}= \delta u_n^t+\prod_{k=-\infty}^{n-1}\frac{u_k^{t+1}}{u_k^t}. 
\label{DKdVnonlinear}
\end{equation} 
We introduce a parameter $\epsilon$ and put $\displaystyle \delta:=e^{-1/\epsilon}$. 
If there exists a one parameter family of solutions to \eqref{DKdVnonlinear} $\displaystyle u_n^t(\epsilon)$,
and if the limit $\displaystyle \lim_{\epsilon \to +0} \epsilon \log u_n^t(\epsilon)=:U_n^t$ exists, 
then, using the identities for arbitrary real numbers $a$ and $b$ 
\begin{align*}
&\lim_{\epsilon \to +0} \epsilon \log \left( e^{a/\epsilon}+e^{b/\epsilon}\right) = \max[a,b],\\
&\lim_{\epsilon \to +0} \epsilon \log \left( e^{a/\epsilon} \cdot e^{b/\epsilon}\right) = a+b,\\
&-\max[-a,-b]=\min[a,b],
\end{align*}
we find that $U_n^t$ satisfies \eqref{UDKDV}.

\begin{figure}[t]
 \begin{center}
 \includegraphics[width=.5\linewidth]{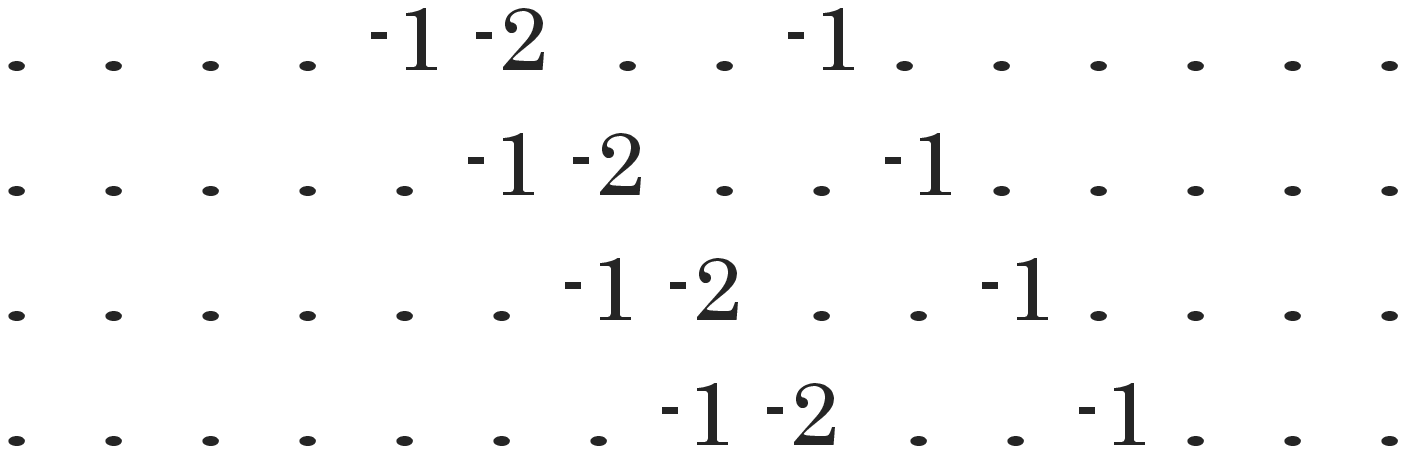} 
 \caption{A time evolution pattern of a background solution. }
 \label{fig03}
 \end{center}
\end{figure}

\begin{figure}[t]
 \begin{center}
 \includegraphics[width=.8\linewidth]{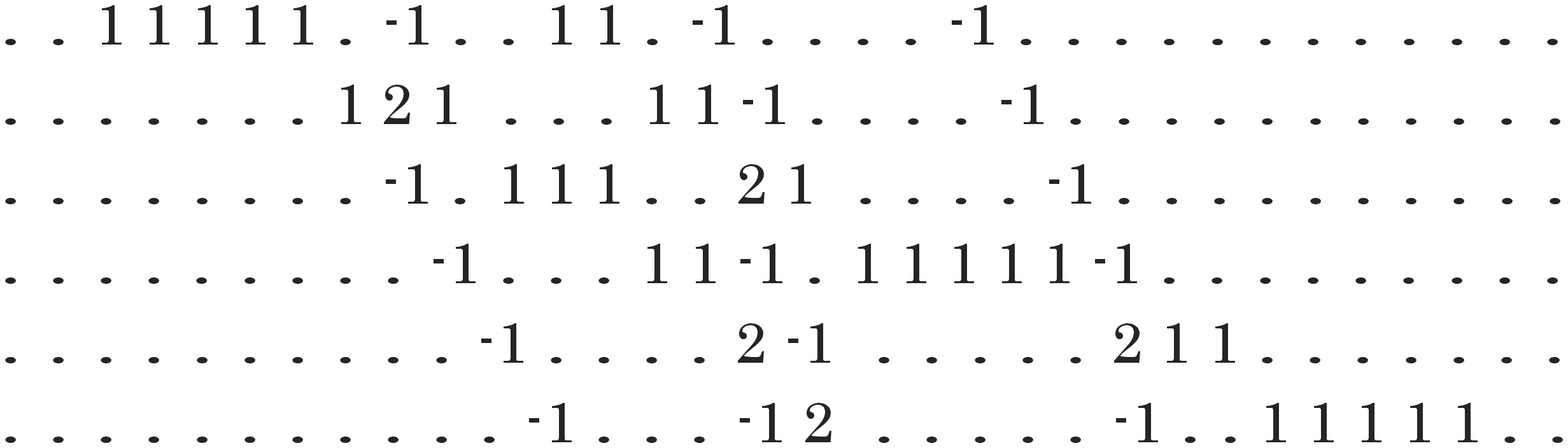} 
 \caption{A time evolution pattern in which negative solitons and ordinary solitons coexist. }
 \label{fig04}
 \end{center}
\end{figure}

Recently, \eqref{UDKDV} has been studied without the constraint of binary values\cite{Hirota, Willoxetal}.
One interesting feature in this extension is that there exist background solutions composed of so-called negative solitons
travelling at speed one\cite{Hirota}.
A typical example of the background solutions is shown in figure~\ref{fig03}.
Furthermore, such negative solitons interact nontrivially with ordinary solitons as shown in figure~\ref{fig04}.
A method to solve the initial value problem for this generalized situation is given in Ref.\cite{Willoxetal}.
From these properties, the dynamical systems described by \eqref{UDKDV} in the general setting are also regarded as integrable cellular automata, just as the BBS.
However, several important features of integrable cellular automata have not been clarified yet, such as the conserved quantities, the relation to integrable lattice models, extension to the case of a cyclic boundary condition and so on, all of which are well understood in the case of BBS.
In this article, we report that a simple transformation reveals the underlying integrable structure of \eqref{UDKDV} for the generalized states.
By this transformation, the relation to the BBS and hence to the integrable lattice models, and generalization to a different boundary condition become apparent.
We explicitly show how to obtain the conserved quantities and give the expression of general solutions to \eqref{UDKDV} where negative solitons and ordinary solitons coexist. 

\section{Conserved quantities for states with negative solitons}
\label{sec2}

We start from the ultradiscretization of the coupled equations \eqref{DKdVc1} and \eqref{DKdVc2}.
By putting $\displaystyle \delta=e^{-1/\epsilon}$ and assuming that the limit
\[
\lim_{\epsilon \to +0} \epsilon\log u_n^t(\epsilon)=: U_n^t,\ \lim_{\epsilon \to +0} \epsilon\log v_n^t(\epsilon)=: V_n^t
\]
exist, we obtain the equations
\begin{align}
U_n^{t+1}&=\min\left[1-U_n^t,V_n^t \right],
\label{CombinatorialR1} \\
V_{n+1}^t&=U_n^t+V_n^t-U_n^{t+1}.
\label{CombinatorialR2}
\end{align}
If we impose the boundary condition $\displaystyle \lim_{|n| \to \infty} U_n^t=\lim_{|n| \to \infty}V_n^t=0$, Eqs. \eqref{CombinatorialR1} and \eqref{CombinatorialR2} are equivalent to \eqref{UDKDV}.
It is easily seen that \eqref{CombinatorialR1} and \eqref{CombinatorialR2} are generalized by introducing a
positive integer constants $L$ and $C$ as
\begin{align}
U_n^{t+1}&=U_n^t+\min\left[L-U_n^t,V_n^t\right]-\min\left[U_n^t,C-V_n^t \right],
\label{CarrierR1} \\
V_{n+1}^t&=U_n^t+V_n^t-U_n^{t+1}.
\label{CarrierR2}
\end{align}
Equations \eqref{CarrierR1} and \eqref{CarrierR2} are known as the ultradiscrete limit of the modified KdV equation
and are interpreted as a BBS of box capacity $L$ with carrier of capacity $C$ \cite{TM, Murataetal}.  
At the same time, they represent the action of the combinatorial R-matrix of an $A_1^{(1)}$ crystal
on a certain symmetric tensor product representation as long as $U_n^t \in \{0,1,\cdots,L\}$ and $ V_n^t \in 
\{0,1,2,\cdots,C\}$\cite{NakayashikiYamada}.
Note that Eqs. \eqref{CarrierR1} and \eqref{CarrierR2} are closed under the values $U_n^t \in \{0,1,\cdots,L\}$ and $ V_n^t \in 
\{0,1,2,\cdots,C\}$.
Hence the time evolution patterns of the BBS are regarded as the ground states of a 2 dimensional integrable lattice model
at temperature 0 and its generalization has been investigated in detail\cite{FOY, HHIKTT}.

When we allow $U_n^t$ or $V_n^t$ in \eqref{CombinatorialR1} and \eqref{CombinatorialR2} to take negative integer values and/or the values greater than $L$ or $C$ respectively, 
they are no longer interpreted as a BBS nor as a realization of the combinatorial R-matrix.
For an arbitrary initial state including negative integer values, let $\displaystyle M:=-\min_{n\in \Z}\left[U_n^0, V_n^0,L-U_n^0,C-V_n^0 \right]$.
When we change variables $\tilde{U}_n^t := U_n^t+M,\ \tilde{V}_n^t:=V_n^t+M,\ \tilde{L}:=L+2M,\ \tilde{C}:=C+2M$, then we find that
Eqs. \eqref{CarrierR1} and \eqref{CarrierR2} are unchanged under this gauge transformation\cite{GilsonNagaiNimmo}, that is,
\begin{align*}
\tilde{U}_n^{t+1}&=\tilde{U}_n^t+\min\left[\tilde{L}-\tilde{U}_n^t,\tilde{V}_n^t\right]-\min\left[\tilde{U}_n^t,\tilde{C}-\tilde{V}_n^t\right],\\
\tilde{V}_{n+1}^t&=\tilde{U}_n^t+\tilde{V}_n^t-\tilde{U}_n^{t+1}.
\end{align*}
Since these equations are closed under the values $\tilde{U}_n^t \in \{0,1,\cdots,L+2M\}$ and $ \tilde{V}_n^t \in 
\{0,1,2,\cdots,C+2M\}$, and initial values $\tilde{U}_n^0,\  \tilde{V}_n^0$ belong to the same sets, we can regard them as
equations for a BBS with box capacity $L+2M$ and carrier capacity $C+2M$, and regard them as a realization of $A_1^{(1)}$ crystal
lattice composed of a larger tensor product representation.
Thus we can conclude that Eqs. \eqref{CarrierR1} and \eqref{CarrierR2} as well as Eqs. \eqref{CombinatorialR1} and \eqref{CombinatorialR2}
do not change their mathematical structure even in the case where the dependent variable take negative integer values.

From the above argument, we find that we can identify the ultradiscrete KdV equation \eqref{UDKDV} as a BBS with larger box capacity
when its dependent variables take negative integer values.
The conserved quantities of the BBS have been widely investigated in general cases and we can apply the results to
this system\cite{FOY,MIT2005}.
In \cite{FOY,Fukuda}, it is shown that the energy functions which are combinatorially determined, are conserved in time, and in
\cite{MIT2005} the path description for the conserved quantities is given.
Here we give another expression of the conserved quantities for \eqref{UDKDV}.

\begin{figure}[t]
 \begin{center}
 \includegraphics[width=1\linewidth]{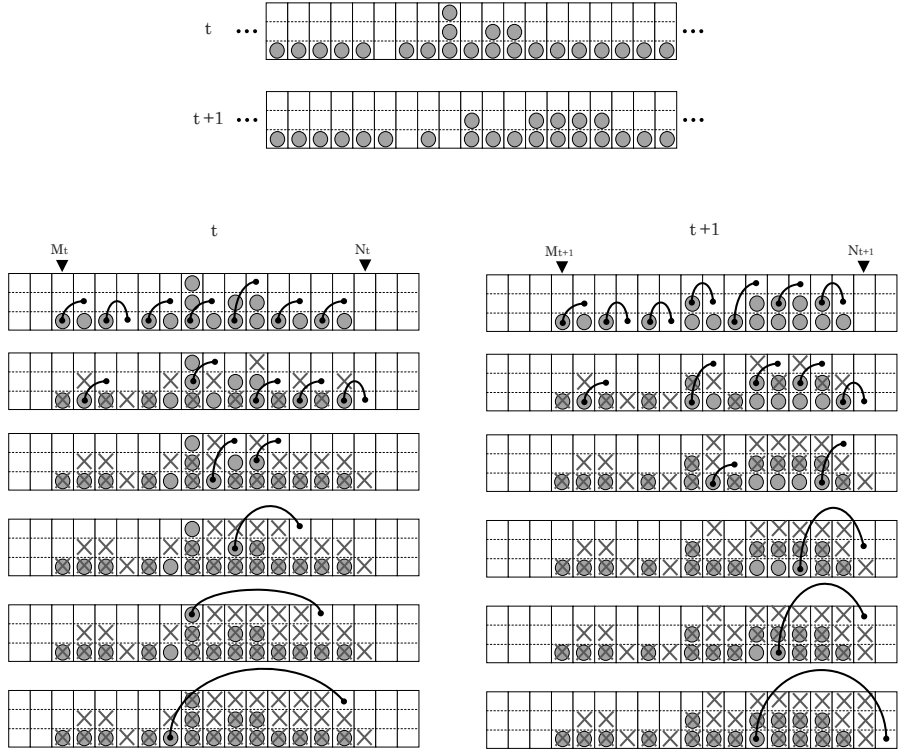} 
 \caption{The scheme to count the conserved quantities of the BBS with larger capacities. We have to choose different sequences of boxes at time step $t$ and $t+1$. From this figure, we have $\tilde{q}_1(t)=7,\ \tilde{q}_2(t)=5,\ \tilde{q}_3(t)=2,\ \tilde{q}_4(t)=1,\ \tilde{q}_5(t)=1,\ \tilde{q}_6(t)=1$
and $\tilde{q}_1(t+1)=7,\ \tilde{q}_2(t+1)=5,\ \tilde{q}_3(t+1)=2,\ \tilde{q}_4(t+1)=1,\ \tilde{q}_5(t+1)=1,\ \tilde{q}_6(t+1)=1$. 
Since $M=1$, $N_t-M_t=14$ and $N_{t+1}-M_{t+1}=14$, we obtain $q_1=7-\frac{14}{2}=0,\ q_2=5-\frac{14}{2}=-2,\ q_3=2,\ q_4=q_5=q_6=1$ in both time steps.}
 \label{fig05}
 \end{center}
\end{figure}

Let $\displaystyle M:=-\min_n[U_n^0,V_n^0,1-U_n^0]$.
Then Eqs. \eqref{CombinatorialR1} and \eqref{CombinatorialR2} are transformed to the BBS with box capacity $1+2M$ (with infinite 
carrier capacity).
Let the number of balls in the $n$th box be $\displaystyle \tilde{U}_n^t$.
We consider the time evolution starting from the time step $t=0$, and
choose the two integers $M_0$ and $N_0$ which satisfy the following conditions.
\begin{enumerate}
\item $U_n^0=U_n^{1}=0$ for $n \le M_0+1$ and $N_0-1 \le n$ ( $\displaystyle \tilde{U}_n^0=\tilde{U}_n^{1}=M$ $(n \le M_0+1,\ N_0-1 \le n)$).
\item $N_0-M_0 \equiv 0$ (mod $2$).
\item No vacancy in the $N_0$th box is connected by an arcline in the first step of the construction of the conserved quantities
which we explain below.
\end{enumerate}
The conditions 2, 3 are used to avoid the redundancy of the definition of conserved quantities. 
For the time step $t$, from the boundary condition, there exists two integers $M_t, \ N_t$ such that $U_n^t=V_n^t=U_n^{t+1}=0$ for $n \le M_t-1$ and $N_t+1 \le n$ ( $\displaystyle \tilde{U}_n^t=\tilde{U}_n^{t+1}=M$ $(n \le M_t+1,\ N_t-1 \le n)$).
Furthermore we impose the following conditions on them.
\begin{align*}
&N_t-M_t \equiv 0 \quad (\mbox{mod $2$}), \\
&M_t-M_0 \equiv t \quad (\mbox{mod $2$}), \\
&N_t-N_0 \equiv t \quad (\mbox{mod $2$}). 
\end{align*}
At each times step $t$, we consider the boxes with indices from $M_t$ to $N_t$.
In other words, we neglect all the balls in the boxes $n < M_t$ and $n > N_t$.
Then, as in the case of the BBS with box capacity one, we draw arclines from balls to vacancies to determine the values of the conserved quantities. (See figure~\ref{fig05}.)
\begin{itemize}
\item Draw an arcline from one of the leftmost balls to one of its nearest right vacancies.
\item Draw an arcline from one of the leftmost balls which are located on the right of the vacancy we have drawn the arcline
to one of its nearest right vacancies.
\item Repeat the above procedure until one has drawn arclines between all the pairs of balls and vacancies.
Note that if we drew an arcline to a vacancy at $n$th box, then the ball from which we draw an arcline in the next step is located
in the box whose index is greater than $n$, and that
there must be no arcline starting from a ball in the last box due to the condition 3.
Let $\tilde{q_1}(t)$ be the number of arclines drawn in this step. 
\item Then neglecting all the boxes and vacancies which are connected by arclines, repeat the above procedure and let $\tilde{q_2}(t)$ be the number of arclines drawn in the second step.
\item Repeat the above procedure until all the balls but those in the $N_t$th box are connected and let $\tilde{q_j}(t)$ ($j=3,4,\ldots$)
be the number of arclines drawn in the $j$th steps. 
\item Let $\DIS q_j:=\tilde{q_j}(t) - \frac{1}{2}(N_t-M_t)\;$ ($j=1,2,...,2M)$ and $q_j:=\tilde{q_j}(t)\;$ $(j=2M+1,2M+2,\ldots)$. 
\end{itemize}
Note that we obtain the updated state of the BBS with carrier capacity $C$ if we exchange the balls with the vacancies connected to them up to the $C$th steps, i.e., if we exchange the first $\DIS \sum_{j=1}^C \tilde{q}_j$ pairs of balls and vacancies.
Then we find the following proposition.
\begin{Proposition}
\label{Prop2-1}
The sequence ($q_1,q_2,q_3,\ldots$) does not depend on $t$.
\end{Proposition}

The statement of the proposition is the direct consequence of the fact that $\displaystyle \sum_{i=1}^k
\tilde{q_i}$ is equal to the number of balls moved by the carrier with capacity $k$ at one time step which is 
the value of $k$th energy function\cite{FOY, HHIKTT}.
Here we map the system to the BBS with box capacity $2M+1$ in which the number of balls is finite, that is, the $n$th box is
vacant for $n \le M_t-1$ and $N_t \le n$.
Of course we can make another choice.
We could impose the condition $N_t-M_t \equiv 1$ (mod $2$) or the condition that a vacancy in the $N_0$th box is connected by an arcline in the first step.
This redundancy is caused by the fact that there are essentially four types of finite systems which are equivalent to this system.
Once we determine the type however, the sequence $(q_1,q_2,\ldots)$ does not depend on $N_t$ and $M_t$ as long as they are sufficiently large and, when there is no negative soliton $q_{2M+j}=\hat{q}_j$ $(j=1,2,\ldots)$, where $\hat{q}_j$ is the $j$th conserved quantity defined in the original BBS.

\section{Soliton solutions interacting with a background state}
\label{sec3}

For the original BBS where every box capacity is one and only a finite number of balls exist, all the solutions are
proved to be constructed by ultradiscretization from the soliton solutions of the discrete KdV equation \eqref{DKdVnonlinear}\cite{MIT2008}. 
Although the ultradiscrete KdV equation \eqref{UDKDV} over arbitrary integers is shown to be equivalent to the BBS with
larger box capacity, the solutions to them can not be the ultradiscrete limit of soliton solutions to \eqref{DKdVnonlinear},
because the boundary condition of the BBS changes such that the number of balls becomes $M$ for $|n| \to \infty$. 
Nevertheless we shall show that solutions can be constructed from the soliton solutions to \eqref{DKdVnonlinear} through ultradiscretization with an appropriate scaling limit.
Here we treate only the case $M=1$.
Our method for constructing the solutions is equally applicable to the case $M \ge 2$, although the analysis of that case is more elaborate.

The coupled equations for the BBS are explicitly written as 
\begin{align}
U_n^{t+1}&=\min\left[3-U_n^t,V_n^t \right],
\label{Oureq1} \\
V_{n+1}^t&=U_n^t+V_n^t-U_n^{t+1},
\label{Oureq2}
\end{align}
with the boundary condition
\begin{equation}
\lim_{|n| \to \infty} U_n^t=\lim_{|n| \to \infty} V_n^t=1.
\label{Ourbc}
\end{equation}
From \eqref{DKdVbilinear} and \eqref{SigmatoUandV}, if we put $\DIS \delta=e^{-L/\epsilon}$ ($L \in \Z_+$) and assume that
\[
\rho_n^t:=\lim_{\epsilon \to +0}\epsilon \log \sigma_n^t
\]
exists, we find 
\begin{equation}
\rho_{n+1}^{t+1}+\rho_n^{t-1}=\max\left[\rho_{n+1}^{t-1}+\rho_n^{t+1}-L, \rho_n^t+\rho_{n+1}^t \right].
\label{UdKdVbilinear}
\end{equation}
From \eqref{SigmatoUandV}, if we put
\begin{align}
U_n^t&=\rho_n^t-\rho_n^{t-1}-\rho_{n+1}^t+\rho_{n+1}^{t-1},
\label{RhotoU}\\
V_n^t&=\rho_{n+1}^t-2\rho_n^t+\rho_{n-1}^t,
\label{RhotoV}
\end{align}
they satisfy \eqref{Oureq1} and \eqref{Oureq2} for $L=3$.
Hence we will construct solutions $\rho_n^t$ of \eqref{UdKdVbilinear} which satisfy the boundary conditions \eqref{Ourbc}.


First let us consider a state where only negative solitons exist.
The corresponding solution is called a background solution \cite{Willoxetal}.
It is a stationary solution of the form $\displaystyle U_n^t=F(n-t)$ where $\displaystyle \;\{F(k)\}_{k=-\infty}^\infty$
is a $01$ sequence with only finite number of $0$s.
Let the number of $0$s be $s+1$ ($s \in \Z_{\ge 0}$), and 
let $b_0=\delta_0=0$, $b_i\in \Z_{\ge 0}$ ($i=1,2,\ldots,s$) and $\delta_i \in\{0,1\}$ $(i=1,2,\ldots,s)$.
We define a $01$ sequence $\displaystyle \{ \tilde{F}(k) \}_{k=-\infty}^\infty$ as
\begin{equation}
\tilde{F}(k)=\left\{
\begin{array}{cl}
0&\;\; k=-\sum_{i=0}^j(2b_i+\delta_i+i) \;(j=0,1,2,\ldots,s),\\
1&\;\;\qquad \mbox{otherwise}.
\end{array}
\right.
\end{equation}
Apparently any background solution $F(k)$ can be expressed as $F(k)=\tilde{F}(k-a)$ for some integers $a,\ s,\ \{b_i\}\ $ and $\{\delta_i\}$.


The $N$ soliton solution to \eqref{DKdVbilinear} is given as
\begin{equation}
\sigma_n^t=1+\sum_{{J \subseteq [N]}\atop{J \ne \phi}}\left( \prod_{i \in J} (-\gamma_i)\left( \frac{1-p_i}{p_i}\right)^t 
\left( \frac{p_i+\delta}{1+\delta-p_i}\right)^n\right)\frac{\prod_{i>j,i,j\in J}(p_i-p_j)^2}{\prod_{i,j \in J}(1-p_i-p_j)},
\end{equation}
where $p_i \in \C$ and $\gamma_i \in \C$ $(i=1,2,\ldots,n)$ are arbitrary but $p_i \ne p_j$ for $i \ne j$\cite{TTMS,HTI}.
We often denote $[k]:=\{1,2,3,\ldots,k\}$ for a positive integer $k$.
To take the ultradiscrete limit, let $\DIS \delta=e^{-L/\epsilon}$, $\DIS p_i =N_i e^{-P_i/\epsilon} $,  
$\DIS -\gamma_i =M_i e^{-C_i/\epsilon} $ $(i=1,2,\ldots,N)$ where $C_i \in \mathbb{R},\  L, P_i, N_i, M_i>0$  and 
$N_i \ne N_j,\ M_i \ne M_j$ for $i \ne j$.
Then taking the limit $\DIS \lim_{\epsilon \to +0} \epsilon \log \sigma_n^t=:\rho_n^t$, we have
\begin{equation}
\rho_n^t=\max\left\{0, \max_{{J \subseteq [N]}\atop{J \ne \phi}}  \left[\sum_{i \in J}(C_i+tP_i-n\min[P_i,L])
-\sum_{{i,j\in J}\atop{i \ne j}} \min[P_i,P_j]   \right]  \right\}.
\label{rhomax}
\end{equation}  

For $L=1$, we find that $U_n^t$ and $V_n^t$ satisfy \eqref{CombinatorialR1} and \eqref{CombinatorialR2}, and hence \eqref{UDKDV}.
In particular,
if $P_i \in \Z_+,\ C_i \in \Z$, \eqref{rhomax} gives an $N$ soliton solution of the original BBS.

We are, however, concerned with Eqs. \eqref{Oureq1} and \eqref{Oureq2} with the boundary condition \eqref{Ourbc}.
Hereafter we put $L=3$ and prepare several symbols and notations:
\begin{align*}
l:&=\sum_{i=1}^s \delta_i, \quad \delta_{s+1}\equiv s-l+1 \;\; (\mbox{mod} \ 2),\quad l':=l+\delta_{s+1}, \\
K:&=l+\sum_{i=1}^sb_i, \quad B_i:=\sum_{j=0}^ib_j,\quad \Delta_i:=\sum_{j=0}^i\delta_i, 
\end{align*} 
We also assume that $\delta_i=1$ for $i=m_1,m_2,\ldots,m_l$ ($1 \le m_1<m_2<\cdots <m_l\le s$), and $m_{l+1}:=s+1$.
Let us consider an $n_{\infty}$ soliton solution $(n_{\infty} \gg 1)$ of the form \eqref{rhomax}.
We assume that $P_i=2$ ($1 \le i \le n_{\infty}-l'$) and $P_i':=P_{i+n_{\infty}-l'}= 1$ $(1 \le i \le l')$.
The phases are chosen as
\begin{align*}
C_i&=\left\{
\begin{array}{cl}
\DIS 1 & (1 \le i \le n_0), \\
\DIS -2j+1 & (n_0+B_{j-1}+1 \le i \le n_0+B_j\;\;(j=1,2,...,s)), \\
\DIS -2s-1 & (n_0+B_s+1 \le i), 
\end{array}
\right.\\
C_i'&\equiv C_{i+n_{\infty}-l'}:=-(m_i-i)\;\;(i=1,2,...,l').
\end{align*}
Then we find 
\begin{align*}
\rho_n^t&=\max_{{0\le k_1 \le n_{\infty}-l'}\atop{0\le k_2 \le l'}}
\left[ \sum_{i=1}^{k_1}(C_i-2(n-t))+\sum_{i=1}^{k_2}(C_i'-(n-t)) \right.\\
&\qquad \qquad  -2k_1(k_1-1)-k_2(k_2-1)-2k_1k_2 \Biggr]\\
&=\max_{{-n_0\le k_1 \le n_{\infty}-l'-n_0}\atop{0\le k_2 \le l'}}
\left[ -(2k_1+k_2)(n-t+2n_0) +{\sum_{i=1}^{k_1}}'\bar{C}_{i}\right.\\
&\qquad \left.+\sum_{i=1}^{k_2}C_i'-2k_1(k_1-1)-k_2(k_2-1)-2k_1k_2 \right]\\
&\qquad \qquad \qquad \qquad \qquad -2n_0(n-t+2n_0)+2n_0^2+3n_0,
\end{align*}
where 
\begin{align*}
\bar{C}_i&:=C_{n_0+i},\\
{\sum_{i=1}^k}' \bar{C}_{i}&:=\left\{
\begin{array}{cl}\DIS
\sum_{i=1}^k \bar{C}_{i}&\quad (k \ge 1), \\
k &\quad (k \le 0).
\end{array}
\right.
\end{align*}
It is obvious that if $\rho_n^t$ is a solution to \eqref{UdKdVbilinear}, $\DIS \rho_{n+b}^{t+c} +at+f(n)$ is 
also a solution for an arbitrary function $f(n)$ and arbitrary values $a,\ b$ and $c$.
Hence we find that $\rho_n^t=\rho_{n_0,n_{\infty}}(n-t)$ is a solution to \eqref{UdKdVbilinear}, where
\begin{align*}
\rho_{n_0,n_{\infty}}(x)&=\max_{{-n_0\le k_1 \le n_{\infty}-l'-n_0}\atop{ 0\le k_2 \le l'}} \\
&\quad \left[ -(2k_1+k_2)x +{\sum_{i=1}^{k_1}}'\bar{C}_{i}+\sum_{i=1}^{k_2}C_i'\right. \\
&\qquad \qquad  -2k_1(k_1-1)-k_2(k_2-1)-2k_1k_2 \Biggr].
\end{align*}
Since $n_0,\ n_{\infty}$ are large but arbitrary positive integers, the limit
\begin{align}
\rho_F(x)&:=\lim_{n_0 \to \infty} \lim_{n_{\infty} \to \infty} \rho_{n_0,n_{\infty}}(x) \notag \\
          &=\max_{{k_1\in \Z}\atop{0\le k_2 \le l'}}
\left[ -(2k_1+k_2)x +{\sum_{i=1}^{k_1}}'\bar{C}_{i}\right. \notag\\
&\quad \left. \qquad +\sum_{i=1}^{k_2}C_i'-2k_1(k_1-1)-k_2(k_2-1)-2k_1k_2 \right]
\label{rhoForiginal}
\end{align}
is also a solution to \eqref{UdKdVbilinear}.

When we rewrite in \eqref{UdKdVbilinear} as
\[
\max_{r\in \Z} \left[ -rx +\max_{{0\le k_2 \le l'}\atop{2k_1+k_2=r}}\left[  {\sum_{i=1}^{k_1}}'\bar{C}_{i}+\cdots \right]\right]
\]
and using the relations such as
\[
a_0+a_2 \ge 2 a_1 \ \longrightarrow \ \max[-2x+a_2,-x+a_1,a_0] = \max[-2x+a_2,a_0]
\]
Eq. \eqref{rhoForiginal} can be simplified as shown in the Proposition \ref{Prop1} below.
First let us define some symbols and notations.
\begin{align*}
C_T&:=\sum_{j=1}^s(2j-1)b_j+\sum_{j=1}^l(m_j-j), \quad Q_0:=C_T+K(K-1)+(K-l)(K-l-1) \\
S_1&:=\{m_1,m_2,\ldots,m_l\},\quad S_0:=[l]\setminus S_1,\quad I :=[2K-l]\setminus \bar{I} \\
\bar{I}&:=\left\{m \in [2K-l]\Big| m=2B_{i-1}+\Delta_{i-1}+2k_i+1,\   k_i=0,1,\ldots,b_{i-1}-1, \ i \in S_0 \right\}
\end{align*}
The quantity $\Theta_r$ is defined for $r \in I$ as
\begin{itemize}
\item If $i \in S_0$ and $\DIS 2B_{i-1}+\Delta_{i-1}+2 \le r \le 2B_i+\Delta_i$, then putting $r=2B_{i-1}+\Delta_{i-1}+2k$
$\ (k \in \mathbb{Z})$, 
\begin{align}
\Theta_r&:=\sum_{j=1}^{i-1}\left[ (2j-1)b_j+(j-\sum_{\mu=1}^j \delta_\mu)\delta_j\right] +(2i-1)k \nonumber \\
&\quad + (B_{i-1}+k+\Delta_{i-1})(B_{i-1}+k+\Delta_{i-1}-1)+(B_{i-1}+k)(B_{i-1}+k-1).
\label{Theta1}
\end{align}
\item If $i \in S_1$, $\DIS 2B_{i-1}+\Delta_{i-1}+1 \le r \le 2B_i+\Delta_i$ and
$r=2B_{i-1}+\Delta_{i-1}+2k$ $\ (k \in \mathbb{Z})$, then $\Theta_r$ is given by \eqref{Theta1}.
\item If $i \in S_1$, $\DIS 2B_{i-1}+\Delta_{i-1}+1 \le r \le 2B_i+\Delta_i$ and 
$r=2B_{i-1}+\Delta_{i-1}+2k+1$ $\ (k \in \mathbb{Z})$, then 
\begin{align}
\Theta_r&:=\sum_{j=1}^{i-1}(2j-1)b_j+\sum_{j=1}^{i}(j-\sum_{\mu=1}^j \delta_\mu) \delta_j+(2i-1)k \nonumber \\
&\quad + (B_{i-1}+k+\Delta_{i})(B_{i-1}+k+\Delta_{i}-1)+(B_{i-1}+k)(B_{i-1}+k-1).
\label{Theta_2}
\end{align}
\end{itemize}

\begin{Proposition}
\label{Prop1}
\begin{equation}
\rho_F(x):=\max\left[\rho_F^{(-)}(x),\rho_F^{(0)}(x),\rho_F^{(+)}(x) \right],
\label{eqrho}
\end{equation}
where
\begin{align}
\rho_F^{(-)}(x)&:=\max_{{k \ge 0}\atop{k\in\mathbb{Z}}}\left[ (2x-1)k-2k(k+1) \right],
\label{rho_minus}\\
\rho_F^{(+)}(x)&:=\left\{
\begin{array}{l}
\DIS\max_{{q \ge 1}\atop{q\in\mathbb{Z}}}\left[ -(2K-l+2q)x-q(2q+4K-2l+2s-1)-Q_0 \right] \quad( \delta_{s+1}=0 ),\\
\DIS \max_{{q \ge 1}\atop{q\in\mathbb{Z}}}\left[ -(2K-l+q)x-(q-1)\left(\frac{q}{2}+2K-l+s\right)-s-Q_0 \right]\quad (\delta_{s+1}=1),
\end{array}
\right.
\label{rho_pluse}\\
\rho_F^{(0)}(x)&:=\max_{{1 \le r \le 2K-l}\atop{r\in\Z}}\left[ \rho^{(r)}(x)  \right],
\label{rho_zero}
\end{align}
and the function $\rho^{(r)}(x)$ is given as
\[
\rho^{(r)}(x):=\left\{
\begin{array}{cl}
-rx-\Theta_r&\quad r \in I, \\
0 &\quad r \notin I.
\end{array}
\right.
\]
\end{Proposition}

\bigskip

\begin{figure}[t]
 \begin{center}
 \includegraphics[width=.8\linewidth]{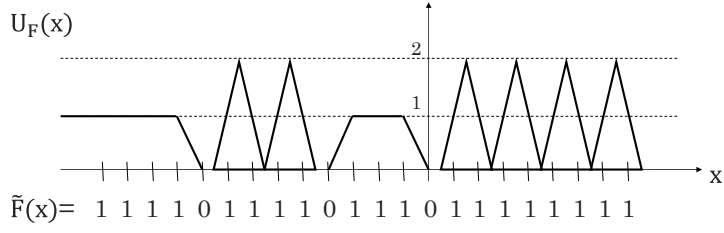} 
 \caption{An example of $\tilde{F}(x)$ and $U_F(x)$. The parameters are $b_1=1,\ \delta_1=1,\ b_2=2,\ \delta_2=0$.}
 \label{fig06}
 \end{center}
\end{figure}

From \eqref{RhotoU}, if we define
\begin{equation}
U_F(x):=\rho_F(x+2)-2\rho_F(x+1)+\rho_F(x),
\label{equf}
\end{equation}
$\DIS U_n^t=U_F(n-t)$ gives a solution to \eqref{Oureq1} and \eqref{Oureq2}.
By using the fact that $\rho_F(x)$ is a continuous and convex piecewise linear function, and  by calculating the intersection points of each linear functions appeared as linear pieces in the function  $\rho_F(x)$,
we obtain the explicit form of $U_F(x)$.
Let us define $t_i(x)$ $(i=1,2,...)$ as
\begin{align*}
t_1(x)&:=\left\{
\begin{array}{cl}
x+2 &\; (-2 \le x \le -1),\\ 
-x &\; (-1 \le x \le 0),\\
0 &\; \mbox{otherwise},
\end{array}
\right.
\\
t_2(x)&:=\left\{
\begin{array}{cl}
2x+5   &\; (-5/2 \le x \le -3/2),\\
 -2x-1  &\; (-3/2 \le x \le -1/2),\\
0   &\; \mbox{otherwise},
\end{array}
\right.\\
t_{2k+1}(x)&:=\sum_{i=0}^{2k}t_1(x+i),\\
t_{2k}(x)&:=\sum_{i=0}^{k-1}t_2(x+2i).
\end{align*}
By straightforward but a little tedious calculation, we find 
\begin{equation}
U_F(x)=U_F^{(-)}(x)+U_F^{(0)}(x)+U_F^{(+)}(x),
\end{equation}
where
\begin{align}
U_F^{(-)}(x)&:=\sum_{k=1}^\infty t_2(x-2k-1), \\
U_F^{(+)}(x)&:=
\left\{
\begin{array}{cl}
\DIS \sum_{k=0}^\infty t_2(x+K+s+2k)&\;\mbox{($\delta_{s+1}=0$)},\\
\DIS \sum_{k=0}^\infty t_1(x+K+s+k)&\;\mbox{($\delta_{s+1}=1$)},
\end{array}
\right. \\
U_F^{(0)}(x)&:=\sum_{i=1}^s t_{2b_i+\delta_i}(x+2B_{i-1}+\Delta_{i-1}+i-1).
\end{align}
An example of the function $U_F(x)$ is shown in figure~\ref{fig06}.
Comparing these expressions with $\tilde{F}(k)$, we find that $\DIS U_F(k)=\tilde{F}(k)$ for $k \in \Z$.
Thus we have proved the following proposition.
\begin{Proposition}
It holds that $U_F(k)=\tilde{F}(k)$ for $k \in \Z$.
\label{Prop2}
\end{Proposition}

\bigskip

Furthermore, comparing $\tilde{F}(k)$ with the conserved quantities $\{ q_j\}$ defined in the previous section, we
have the following proposition.
\begin{Proposition}
\label{conservation}
The number of negative solitons is equal to $-q_1-q_2$, i.e., $s+1=-q_1-q_2$, and 
$q_1-q_2$ is the number of solitons with amplitude $1$ which is used to construct the
background solution, i.e., $q_1-q_2=l'$.
\end{Proposition}


\bigskip

Next we consider the solutions in which $m$ ordinary solitons exist in a background state.
Since $N$ soliton solutions are given by \eqref{rhomax} and the parameters for the background solution are already known, 
we have only to add new parameters $\DIS \{\hat{P}_i\}_{i=1}^m$ $({}^\forall i \ \hat{P}_i \ge 3)$ and $\DIS \{\hat{C}_i\}_{i=1}^m$.
When we denote the set of indices for a background solution $\mathcal{B}$, then the solution is given by
\begin{align*}
\rho_n^t&=\max_{J_1 \subseteq [m], J_2 \subseteq \mathcal{B}}\left[\sum_{i \in J_1}(-\min[3,\hat{P}_i]n+\hat{P}_it+\hat{C}_i) + 
 \sum_{i \in J_2}(-P_i(n-t)+C_i) \right.\\
&\qquad -\sum_{i\ne j, i,j \in J_1}\min[\hat{P}_i,\hat{P}_j] -\sum_{i\ne j, i,j \in J_2}\min[P_i,P_j] 
-2 \sum_{i \in J_1, j\in J_2} \min[\hat{P}_i,P_j]\Biggr]\\
&=\max_{J_1 \subseteq [m], J_2 \subseteq \mathcal{B}}\left[\sum_{i \in J_1}(-3n+\hat{P}_it+\hat{C}_i) + 
 \sum_{i \in J_2}(-P_i(n-t)+C_i) \right.\\
&\qquad -\sum_{i\ne j, i,j \in J_1}\min[\hat{P}_i,\hat{P}_j] -\sum_{i\ne j, i,j \in J_2}\min[P_i,P_j] 
-2 \sum_{ j\in J_2} |J_1|P_j\Biggr]\\
&=\max_{J_1 \subseteq [m]}\left[ \sum_{i \in J_1}(-3n+\hat{P}_it+\hat{C}_i)   -\sum_{i\ne j, i,j \in J_1}\min[\hat{P}_i,\hat{P}_j] \right.\\
&\qquad+\max_{J_2 \subseteq \mathcal{B}}\left.\left[ \sum_{i \in J_2}(-P_i(n-t+2|J_1|)+C_i) -\sum_{i\ne j, i,j \in J_2}\min[P_i,P_j]  \right]
\right].  
\end{align*}
Here we used the abbreviation
\[
\max_{J \subseteq \mathcal{A}}\left[\cdots \right] \equiv \max\left\{ 0, \max_{J \subseteq \mathcal{A}, J \ne \phi}\left[\cdots \right]\right\}.
\]
Therefore we obtain the following theorem.
\begin{Theorem}
\label{Th1}
The solution to \eqref{UdKdVbilinear} representing the interaction of $m$ soliton solutions with the background given by 
\eqref{eqrho}, has the expression:
\begin{equation}
\rho_n^t=\max_{J \subseteq [m]}\left[ \sum_{i \in J}(-3n+\hat{P}_it+\hat{C}_i)  -\sum_{{i\ne j}\atop{i,j \in J}}\min[\hat{P}_i,\hat{P}_j]  +\rho_F(n-t+2|J|)  \right].
\label{Lastsolution}
\end{equation}
\end{Theorem}
The multi-soliton solutions with a background is also given by Y. Nakata in an algebraic form (nested maximization) 
with the ultaradiscrete vertex operators\cite{Nakata}.
So far we do not find the exact correspondence between our solution \eqref{Lastsolution} and his formula.

\medskip

\begin{figure}[t]
 \begin{center}
 \includegraphics[width=.8\linewidth]{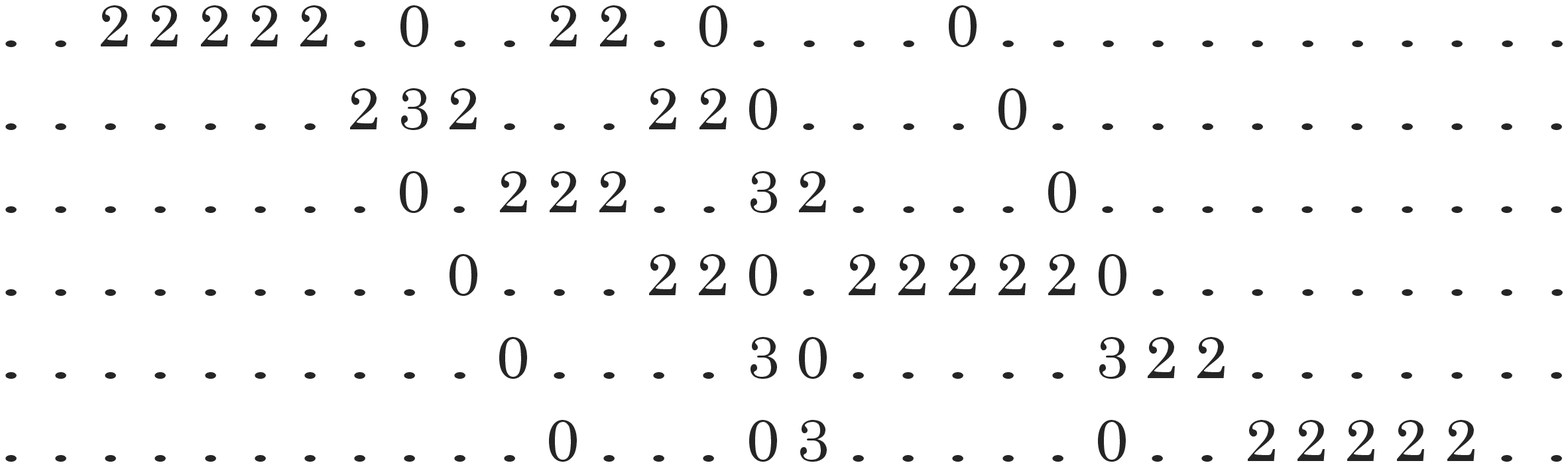} 
 \caption{A solution $U_n^t$ given by \eqref{Lastsolution}.  Here $U_n^t=\rho_n^t-\rho_n^{t-1}-\rho_{n+1}^t+\rho_{n+1}^{t-1}$.
The symbol `.' denotes `1'.
 The parameters are $\hat{P}_1=4,\ \hat{C}_1=-7,\ \hat{P}_2=7,\ \hat{C}_2=-5,\ b_1=2,\ \delta_1=0,\ b_2=1,\ \delta_2=1$. 
When we make gage transformation as $U_n^t-1 \to U_n^t$, this time evolution pattern coincides with that of figure~\ref{fig04}.}
 \label{fig07}
 \end{center}
\end{figure}

From \eqref{Lastsolution}, we find that the phase shift of negative solitons after collision of an ordinary soliton is $-2$. 
An example of a time evolution pattern is shown in figure~\ref{fig07}.
Since the solution \eqref{Lastsolution} shows that the ordinary solitons move freely for $t \gg 1$ as those in the BBS with box capacity one, comparing with the construction
of conserved quantities, we find the following proposition\cite{YYT}.
\begin{Proposition}
For a solution to \eqref{UDKDV} determined by \eqref{Lastsolution},
let $Y$ be the Young diagram corresponding to the partition $(q_3,q_4,q_5,\cdots)$ constituted by the conserved quantities,
that is, the length of its $i$th column of $Y$ is $q_{i+2}$.
Then, the length of $j$th row in $Y$ is equal to $\hat{P}_j-2$.
\end{Proposition} 
%
%
\section{Concluding remarks}
We constructed the conserved quantities of the ultradiscrete KdV equation \eqref{UDKDV} with negative solitons by using the transformation
to the BBS with larger box capacities.
The solutions to \eqref{UDKDV} in which negative solitons and ordinary solitons coexist are also obtained as the solutions to this extended BBS.
In the BBS with box capacity one, every state is obtained by ultradiscretization of a soliton solution of the discrete KdV equation,
and using this fact, initial value problem is solved with elementary combinatorial methods.
We conjecture that this fact also holds for the extended BBS and can be used to solve the initial value problems.
Investigating this conjecture is a problem we wish to address in the future.

\section*{Acknowledgement}
The authors wish to thank Professors Atsushi Nagai and Ralph Willox for useful discussions and comments.


\end{document}